\documentstyle[12pt]{article}

\catcode`\@=11
\long\def\@makefntext#1{
\protect\noindent \hbox to 3.2pt {\hskip-.9pt
$^{{\ninerm\@thefnmark}}$\hfil}#1\hfill}		

 \def\@makefnmark{\hbox to 0pt{$^{\@thefnmark}$\hss}}  

\def\ps@myheadings{\let\@mkboth\@gobbletwo
\def\@oddhead{\hbox{}
\rightmark\hfil\ninerm\thepage}
\def\@oddfoot{}\def\@evenhead{\ninerm\thepage\hfil
\leftmark\hbox{}}\def\@evenfoot{}
\def\sectionmark##1{}\def\subsectionmark##1{}}

\newcounter{sectionc}\newcounter{subsectionc}\newcounter{subsubsectionc}
\renewcommand{\section}[1] {\vspace{0.6cm}\addtocounter{sectionc}{1}
\setcounter{subsectionc}{0}\setcounter{subsubsectionc}{0}\noindent
	{\bf\thesectionc. #1}\par\vspace{0.4cm}}
\renewcommand{\subsection}[1] {\vspace{0.6cm}\addtocounter{subsectionc}{1}
	\setcounter{subsubsectionc}{0}\noindent
	{\it\thesectionc.\thesubsectionc. #1}\par\vspace{0.4cm}}
\renewcommand{\subsubsection}[1]
{\vspace{0.6cm}\addtocounter{subsubsectionc}{1}
	\noindent {\rm\thesectionc.\thesubsectionc.\thesubsubsectionc.
	#1}\par\vspace{0.4cm}}
\newcommand{\nonumsection}[1] {\vspace{0.6cm}\noindent{\bf #1}
	\par\vspace{0.4cm}}

\newcounter{appendixc}
\newcounter{subappendixc}[appendixc]
\newcounter{subsubappendixc}[subappendixc]

\renewcommand{\appendix}[1] {\vspace{0.6cm}
        \refstepcounter{appendixc}
        \setcounter{figure}{0}
        \setcounter{table}{0}
        \setcounter{equation}{0}
        \renewcommand{\thefigure}{\Alph{appendixc}.\arabic{figure}}
        \renewcommand{\thetable}{\Alph{appendixc}.\arabic{table}}
        \renewcommand{\theappendixc}{\Alph{appendixc}}
        \renewcommand{\theequation}{\Alph{appendixc}.\arabic{equation}}
        \noindent{\bf Appendix \theappendixc #1}\par\vspace{0.4cm}}

\def\abstracts#1{{
	\centering{\begin{minipage}{30pc}\tenrm\baselineskip=12pt\noindent
	\centerline{\tenrm ABSTRACT}\vspace{0.3cm}
	\parindent=0pt #1
	\end{minipage}}\par}}

\renewenvironment{thebibliography}[1]
	{\begin{list}{\arabic{enumi}.}
	{\usecounter{enumi}\setlength{\parsep}{0pt}
\setlength{\leftmargin 1.25cm}{\rightmargin 0pt}
	 \setlength{\itemsep}{0pt} \settowidth
	{\labelwidth}{#1.}\sloppy}}{\end{list}}

\newcounter{itemlistc}
\newcounter{romanlistc}
\newcounter{alphlistc}
\newcounter{arabiclistc}

\newcommand{\fcaption}[1]{
        \refstepcounter{figure}
        \setbox\@tempboxa = \hbox{\tenrm Fig.~\thefigure. #1}
        \ifdim \wd\@tempboxa > 6in
           {\begin{center}
        \parbox{6in}{\tenrm\baselineskip=12pt Fig.~\thefigure. #1}
            \end{center}}
        \else
             {\begin{center}
             {\tenrm Fig.~\thefigure. #1}
              \end{center}}
        \fi}

\newcommand{\tcaption}[1]{
        \refstepcounter{table}
        \setbox\@tempboxa = \hbox{\tenrm Table~\thetable. #1}
        \ifdim \wd\@tempboxa > 6in
           {\begin{center}
        \parbox{6in}{\tenrm\baselineskip=12pt Table~\thetable. #1}
            \end{center}}
        \else
             {\begin{center}
             {\tenrm Table~\thetable. #1}
              \end{center}}
        \fi}

\def\@citex[#1]#2{\if@filesw\immediate\write\@auxout
	{\string\citation{#2}}\fi
\def\@citea{}\@cite{\@for\@citeb:=#2\do
	{\@citea\def\@citea{,}\@ifundefined
	{b@\@citeb}{{\bf ?}\@warning
	{Citation `\@citeb' on page \thepage \space undefined}}
	{\csname b@\@citeb\endcsname}}}{#1}}

\newif\if@cghi
\def\cite{\@cghitrue\@ifnextchar [{\@tempswatrue
	\@citex}{\@tempswafalse\@citex[]}}
\def\citelow{\@cghifalse\@ifnextchar [{\@tempswatrue
	\@citex}{\@tempswafalse\@citex[]}}
\def\@cite#1#2{{$\null^{#1}$\if@tempswa\typeout
	{IJCGA warning: optional citation argument
	ignored: `#2'} \fi}}

\def\fnt#1#2{\footnotetext{\kern-.3em
	{$^{\mbox{\sevenrm #1}}$}{#2}}}

 1
 1
 1

\font\tenbf=cmbx10
\font\tenrm=cmr10
\font\tenit=cmti10

\font\ninerm=cmr9

\newcommand{\nn}{\noindent}
\newcommand{\non}{\nonumber}

\newcommand{\epem}{e^+e^-}
\newcommand{\ee}{e^+e^-}
\newcommand{\ra}{\rightarrow }
\newcommand{\lra}{\longrightarrow }
\newcommand{\SM}{{\cal SM}}
\newcommand{\THDM}{2{\cal HDM}}
\newcommand{\MSSM}{{\cal MSSM}}
\newcommand{\SUSY}{{ SUSY}}
\newcommand{\CP}{{\cal CP}}
\newcommand{\beq}{\begin{eqnarray}}
\newcommand{\eeq}{\end{eqnarray}}
\newcommand{\ga}{\gamma \gamma}
\newcommand{\tb}{{\rm tan} \beta}
\newcommand{\tg}{{\rm tan} \beta}

\begin{document}

\centerline{\tenbf HIGGS SEARCH AT $e^+e^-$ AND $\gamma\gamma$}
\baselineskip=16pt
\centerline{\tenbf COLLIDERS\footnote{To appear in Proc. Budapest
Workshop on Electroweak Symmetry Breaking, Budapest, July 11--13, 1994.}}
\vspace{0.8cm}
\centerline{\tenrm JAN KALINOWSKI\footnote{
Supported by the Polish Committee for Scientific Research Grant 2 P302 095 05}}
\baselineskip=13pt
\centerline{\tenit Institute of Theoretical Physics, Warsaw University }
\baselineskip=12pt
\centerline{\tenit Ho\.za 69, 0068 Warsaw, Poland}
\vspace{0.3cm}
\centerline{\tenrm }
\vspace{0.9cm}
\abstracts{\nn The prospects for discovering Higgs particles and studying
their fundamental properties at future high--energy electron--positron and
photon-photon colliders are reviewed. Both the Standard Model Higgs boson
and the Higgs particles of its minimal supersymmetric extension are discussed.
We also comment on a two Higgs doublet model searches at LEP 1.
We update various results by taking into account the  value of the top
quark mass obtained by the CDF Collaboration and by including radiative
corrections.}{}{}

\vfil
\rm\baselineskip=14pt
\section{Introduction}
The Higgs mechanism of electroweak  symmetry breaking is a
crucial ingredient of the Standard Model ($\SM$). It allows to
give masses to weak gauge bosons without affecting the
renormalizability of the model.  At the same time fermions get
their masses via  Yukawa interaction with the ground state of
the Higgs field.  This goal is achieved by employing one
iso-doublet of scalar Higgs fields which leads to one physical
Higgs boson.\cite{Gunion}  The fundamental nature of
mass generation of quarks, leptons and weak gauge bosons
requires this picture to be verified experimentally in all its
aspects. Therefore the Higgs search constitues one of the  most
important physics goals at current and future experiments and is
one of the motivations for building new
accelerators.\cite{LHC,EE1,EE2,EE3}  If the
Higgs particle is found its properties must be measured, in
particular the couplings to other particles which are uniquely
fixed by Higgs mechanism.

Although one can argue that it is a matter of time until the
Higgs boson of the $\SM$ is found, some theoretical problems
related to its existence suggest that it is necessary to look
beyond the Standard Model.  Scalar fields have a nice property
that thay can have a nonzero vacuum expectation value without
breaking Lotentz invariance and thus can trigger breakdown of
gauge symmetry. However they also have a bad property  of
aquiring  quadratic divergences through radiative corrections.
The correction to the mass of the Higgs boson is $\delta m^2
\sim g^2 \Lambda^2$, where $\Lambda$ is a physical scale beyond
which the low energy theory no longer applies. To understand the
physics at the Fermi scale $\sim 10^2$ GeV it would be
inappropriate to have $\Lambda$ of the order of the unification
scale $\sim 10^{15}$ GeV or the Planck scale $\sim 10^{19}$ GeV
but rather in the TeV region. This is so called the naturalness
or hierarchy  problem.  Therefore we are led to
consider means to stabilize the Higgs sector below or at TeV
scale. The only way to protect masses of elementary scalar
particles is supersymmetry.\cite{Haag} This symmetry relates
bosons to fermions and  makes the bosons to behave as well
as fermions.

In this review we start in section 2 with the presentation of
the  Higgs sector in $\SM$. Various decay modes of the Higgs
boson and its production mechanisms at current and future
$\epem$ and $\ga$ colliders with an energy in the range 300 -
500 GeV are discussed.  In section 3 we  discuss the  minimal
supersymmetric ($\MSSM$) extension of the model in which two
Higgs doublets are present. Although some features are specific
to this version of the model, the  pattern of phenomena to be
discussed is characteristic to more general SUSY models.  We
will also comment on Higgs search in a two Higgs doublet model
($\THDM$) at LEP 1.  Section 4 contains conclusions and outlook.

We  discuss Higgs searches in $\ee$ and $\ga$
collisions.  The Higgs search at hadron colliders has been
discussed in the talk of Poggioli.\cite{Poggioli} We update various
results for Higgs masses and couplings, partial decay widths and
production cross sections  taking into account the value of the
top quark mass $m_t=175$ GeV consistent with the value recently
published by the CDF collaboration\cite{CDFTOP} and with the
favored value obtained from a global fit\cite{LEPTOP} of
electroweak precision measurements at LEP and SLC. Sometimes we
will take the values 150 or 200 GeV which can be viewed as
conservative lower and upper bounds on the top mass,
respectively.

\section{Search for the Higgs boson of the Standard Model}
\vspace{-0.7cm}
\subsection{The Higgs sector of the $\SM$ }
\vspace{-0.35cm}
In the Standard Model the  mass of the Higgs boson is  the only
unknown parameter.  Once the  mass is fixed,  the profile of the
Higgs particle can be predicted completely.

Even though the value of the Higgs mass cannot be predicted,
interesting constraints can be derived from assumptions: a) on
the energy range within which the model is valid before
perturbation theory breaks down at a scale $\Lambda$ and new
dynamical phenomena would emerge, and  b) stability of the
vacuum.\\
a)  Since the strength of the Higgs self--interaction is
determined by the Higgs mass itself $M_H$, the condition $M_H<
\Lambda$ sets an upper limit on the Higgs mass in the Standard
Model. Thorough   analyses lead to an estimate of about 630 GeV
for the upper limit of $M_H$. On the other hand, if
the Higgs mass is less than 180 to 200 GeV, the Standard Model
can be extended up to the GUT scale $\Lambda_{{\rm GUT}} \sim
10^{15}$ GeV with weakly interacting particles. Including the
effect of $t$--quark loops on the running coupling, a detailed
analysis predicts the area of the allowed $(m_t, M_H)$ values
shown in Fig.~1 for several values of the cut--off parameter
$\Lambda$.\cite{Lind}\\
\nn b) Quantum corrections due to top-quark loops to the quartic
Higgs coupling are negative, driving the coupling to negative
values for which the vacuum becomes unstable.  For top masses
larger than about 100 GeV, this leads to a lower limit on the
Higgs mass,  Fig.~1.

 From the above arguments the Higgs mass could well be expected in the
window $100 < M_H < 180$ GeV for a top mass value of 150 GeV and
$160 < M_H <200$ GeV for $m_t\sim 175$ GeV. The mass range $M_Z
- 2M_Z$ is usually referred to as intermediate Higgs mass range.

\vspace{1cm}
\begin{small}
\nn \hspace*{10cm} \begin{minipage}[r]{45mm}{Fig.~1 Values of
the top and Higgs masses for which the Standard Model can be
extended up to the scale
$\Lambda$; from Ref.\cite{Lind} The lower bound is
derived from vacuum stability. The dashed line shows the upper
limit for the mass og the lighest scalar Higgs boson of $\MSSM$, see Section
3 and Fig.~6} \end{minipage}
\end{small}

\vspace{4cm}
\subsection{Decay modes}
\vspace{-0.35cm}
  The strength of the couplings of the Higgs boson to fermions and to
the electroweak gauge bosons $V=W,Z$ is set by their masses:
\beq
g_{ffH} & =& \left[ \sqrt{2} G_F \right]^{1/2} m_f \\
g_{VVH} & =& 2 \left[ \sqrt{2} G_F \right]^{1/2} M_V
\eeq

\nn In the Born approximation the width of the Higgs decay into
fermion pairs is
\beq
\Gamma (H \ra f\bar{f})= N_c\frac{G_F m_f^2}{4 \sqrt{2} \pi} \ M_H \
\beta^3 \label{Hll}
\eeq
with $\beta=(1- 4m_f^2/M_H^2)^{1/2}$, $N_c=1$ for leptons and
$N_c=3$ for quarks. For quark pairs, one has to use running
quark masses evaluated at the scale $\mu=M_H$\cite{V4} which
take the bulk of QCD corrections. For example, in the case of
the $b$ quark and for Higgs masses around 100 GeV, it amounts to
a reduction of the $ H \ra b\bar{b}$ decay width by more than 50\%.

Above the $H \ra WW$ and $ZZ$ decay thresholds, the partial width into
massive gauge boson pairs may be written as\cite{Lee}
\begin{eqnarray}
\Gamma (H \ra VV) = \delta_V \frac{\sqrt{2}G_F}{32 \pi} M_H^3 (1-4x+12x^2)
\beta\,,
\end{eqnarray}
where $x=M_V^2/M_H^2$, $\beta=\sqrt{1-4x}\,$ and $\delta_V=2(1)$ for $V=W(Z)$.

Below the threshold for two real bosons, the Higgs particle can decay into
real and virtual $VV^*$ pairs (primarily $WW^*$ pairs above $M_H
\sim 110$ GeV) followed by $V^*$ decay into a fermion pair.
The partial decay width is given by\cite{Keung}
\begin{eqnarray}
\Gamma (H \ra VV^*) = \frac{3 G_F^2 M_V^4}{16 \pi^3} M_H R(x) \delta_V'\,,
\end{eqnarray}
with $\delta'_W=1$ and $\delta_Z' =7/12-10\sin^2\theta_W/9+40\sin^4\theta_W/27$
and
\begin{eqnarray}
R(x) =  \frac{3-24x+60x^2}{(4x-1)^{1/2}} \arccos \frac{3x-1}
{2x^{3/2}} -\frac{1-x}{2x} (2-13x+47x^2)
- (\frac{3}{2}-9x+6x^2) \log x \nonumber
\end{eqnarray}

\vspace{1cm}
\begin{small}
\hspace*{10cm}
\begin{minipage}[r]{45mm}
{ Fig.~2 Total decay width (a) and decay branching ratios (b) of the
$\SM$ Higgs boson; the top quark mass is fixed to $m_t=175$ GeV. The QCD
corrections to the hadronic decay modes are included.}
\end{minipage}
\end{small}

\newpage
By adding up all possible decay channels we obtain the total width shown
in Fig.~2a for $m_t=175$ GeV. The Higgs particle is
very narrow ($\Gamma(H) \leq 10$ MeV) below the (virtual) gauge boson
channels and becomes rapidly wider reaching $\sim$ 1 GeV at the $ZZ$
threshold.  Only above $M_H \geq 250$ GeV it becomes wide enough
to be resolved experimentally.

The branching ratios of the main decay modes are displayed in
Fig.~2b from Ref.\cite{Abdel}, an update of Ref.\cite{Kleiss}. A
large variety of channels will be accessible for Higgs masses
below 140 GeV. By far the dominant mode are $b \bar{b}$ decays,
yet $c \bar{c}$, $\tau^+ \tau^-$ and $gg$ still occur at a level
of several percent.  The branching ratios for the $H \ra \ga$
and $\gamma Z$ are small, being of ${\cal O}(10^{-3})$.  Above
the mass value $M_H=140$ GeV, the Higgs boson decay into $W$'s
becomes dominant, overwhelming all other channels once the decay
mode into two real $W$'s is kinematically possible.

Although the coupling $H\ga$ is small, it is very interesting
since it is sensitive to all charged particles in the loop and
can be used as a possible probe of new particles whose masses
are generated by the Higgs mechanism. This will be important
when we will  discuss Higgs production in $\ga$ collisions.

\subsection{\SM\ Higgs search at LEP}
\vspace{-0.35cm}
Prior to LEP only the Higgs masses up to $\sim$ 5 GeV have been excluded.
The most comprehensive search for Higgs particles has been carried out
in $Z$ decays at LEP,\cite{LEPHiggs} based on the Bjorken
process $Z \rightarrow
Z^*H$. Both visible $Z^*\ra q\bar{q}$ and $l^+l^-$ and invisible
$Z^*\ra \nu\bar{\nu}$ decay modes of the virtual $Z$ in
processes with
Higgs boson decays into $b\bar{b}$, $c\bar{c}$ and
$\tau^+\tau^-$ have been looked for.
By now, a lower bound on the Higgs mass of about $M_H \geq 63.5$ GeV
can be established. This limit can be raised by a few GeV by
accumulating more statistics. At $M_H\sim 70$ GeV another
production process $Z\ra H\gamma$ can also be explored.
In the second phase of LEP with a total energy
close to 200 GeV Higgs particles can be searched for up to
masses of $\sim 110$ GeV in Higgs bremsstrahlung off the $Z$
line.  In the range $m_Z\pm 10$ GeV the sensitivity
will be weaker due to irreducible background coming from $\epem
\ra ZZ$ and the $b$ quark tagging will be necessary to discover
or exclude the Higgs. Higher energy colliders
are required to sweep the entire mass range for the Higgs particle.

\subsection{Production mechanisms at future $\epem$  colliders}
\vspace{-0.35cm}
 At $\ee$ linear colliders operating in the 300--500 GeV energy range, the
main production mechanisms for Higgs particles are the following
processes:\cite{EE2}
\begin{eqnarray}
{\rm bremsstrahlung \ process} : & & \ee \lra (Z) \lra Z+H
\label{brem} \\
{\rm WW \ fusion \ process} : & & \ee \lra \bar{\nu}\ \nu \ (WW)\lra
\bar{\nu}\ \nu \ +H \label{wfus} \\
{\rm ZZ \ fusion \ process} : & & \ee \lra e^+ e^- (ZZ) \lra e^+ e^- +
H \label{zfus}\\
{\rm radiation \ off \ top}:& & \ee \lra (Z, \gamma) \lra (t \bar{t})
 \lra t \bar{t} + H \label{offt}
\end{eqnarray} \nopagebreak
The mass dependence of the \nopagebreak cross sections at
$\sqrt{s}=500$ GeV is shown in Fig.~3.

For $M_H$ in the range $150 - 200$ GeV  the cross
sections for the bremsstrahlung and the $WW$ fusion processes
are of comparable size at $\sqrt{s}=500$ GeV, while the $ZZ$
fusion cross section is smaller by an order of magnitude. With
$\sigma \sim $ 100 fb, a total of $\sim $ 2000 Higgs
particles per year can be created at an
integrated luminosity of $\int {\cal L} =20\ {\rm fb}^{-1}$. For
Higgs masses below 100 GeV, the cross section for Higgs
radiation off top quarks Eq.~(\ref{offt})  is of the order of a
few fb;\cite{X8} this process can be used only to measure the
$t\bar{t}H$ Yukawa coupling once the
Higgs boson is detected in the previous processes.

\vspace*{1cm}
\begin{small}
\hspace*{10cm}\begin{minipage}[r]{45mm}
{ Fig.~3 Production cross sections for $\SM$ Higgs particles
at $\sqrt{s}=500$ GeV.}
\end{minipage}\end{small}

\vspace*{5cm}

\vspace{3mm}
\nn {\it 2.4.1. Higgs Bremsstrahlung}

The cross section for the process Eq.~(\ref{brem}) can be
written in a compact form
\beq
\sigma(\ee \ra ZH) = \frac{G_F^2 M_Z^4}{96 \pi s} (v_e^2+a_e^2)
\ \lambda^{1/2} \frac{ \lambda+ 12M_Z^2/s}{(1-M_Z^2/s)^2} \label{sigmaZH}
\eeq
where $a_e=-1$ and $v_e=-1+4s_W^2$ are the $Z$ charges of the electron and
$\lambda=(1-M_H^2/s-M_Z^2/s)^2-4M_H^2M_Z^2/s^2$ is the usual two--particle
phase
space function.

Due to kinematical constraints familiar from low-energy $\ee$
experiments there are several strategies  that
can be adopted. First, the recoiling $Z$
boson in the two--body reaction $\ee \ra ZH$ is
mono--energetic  and the mass of the Higgs
boson can be derived from the energy of the $Z$ boson, $M_H^2 =s -2\sqrt{s} E_Z
+M_Z^2$. Second, the Higgs boson can be reconstructed from its
decay products with an additional constraint provided
by the recoiling system which should have an invariant mass to
be equal $M_Z$. Finally full reconstruction of the events can be
attempted.

In the first two cases  the initial $e^+$ and $e^-$ beam
energies  have to be well known. However,
beamstrahlung smears out the c.m. energy and the system moves along the beam
axes.  The intensity of the beamstrahlung depends on the machine
design. In this context promising results\cite{E6} have been obtained in
the DESY--Darmstadt and the
TESLA design studies. For these designs  the smearing of
the missing mass is expected to be of  the same magnitude as the
experimental uncertainties in
the reconstruction of the $Z$ boson in the leptonic decay channels.

Since the recoiling $Z$ boson remains approximately mono--energetic,
even if beamstrahlung is taken into account, it is easy to separate the
signal from the background\cite{PR4}. Only for  Higgs masses close to
$Z$ mass the selection of $b\bar{b}$
final states from $H$ decays by means of flavor tagging through
vertex detection will be necessary in order to eliminate the
dominant background from  double $Z$--production
$\ee \ra ZZ$. For example, at $\sqrt{s}=500$ GeV and integrated luminosity
$\int{\cal L}=20$ fb$^{-1}$ one expects 50 ZH and 142 ZZ events
without b-tagging, and 38 and 33 with b-tagging, respectively.
 For masses $M_H$ between 100 and 160 GeV the
dominant background  from
single $Z$--production in $\ee \ra Z Z^*(\ra q \bar{q})$ and
$\ee \ra Z+WW^*(\ra q
\bar{q}')$ is suppressed by at least one
power of the electroweak coupling relative to the signal.
Beyond 160 GeV and 180 GeV, the reactions with three gauge
bosons in the final state, $\ee \ra Z+WW$ and $\ee \ra Z+ZZ$ are the main
background channel with the invariant mass of the $WW$ or $ZZ$
final states  broad as opposed to the resonance structure of the signal
and missing mass technique can be used
up to kinematical limit.

An interesting feature of the bremsstrahlung process is that
the angular distribution of the $Z/H$ bosons is
sensitive to the spin of the Higgs particle.\cite{Barger}
At high energies the $Z$ boson in Eq.~(\ref{brem}) is produced
with longitudinal
polarization and the angular distribution approaches the spin--zero
angular distribution asymptotically  ${\rm d}\sigma/{\rm d}\cos\theta
\ra \frac{3}{4} \sin^2 \theta $.
For a pseudoscalar state $A(0^{-+})$  (realized for example in
$\THDM$ and $\SUSY$ models) the effective point-like
coupling $ZZA$ is a
P--wave coupling, as opposed to
a S--wave for a scalar Higgs, and in this case the angular
distribution is  d$\sigma(ZA)/$dcos$\theta \sim
1- \frac{1}{2}\sin^2\theta$, independent of the energy.
The angular distributions specific to  Higgs production
in $\ee \ra ZH$ or $\ee \ra ZA$ are different from  the process $\ee
\ra ZZ$ which is mediated by electron exchange in the
$t$--channel and the amplitude is built up by many partial
waves, peaking in the
forward/backward direction. The three distributions (ZH, ZA, ZZ)
are compared with each other
in Fig.~4, which demonstrates the specific character of the Higgs production
process.

Since the $Z$ bosons from $ZH$ production are expected to be asymptotically
polarized longitudinally (by contrast, the $Z$ bosons
from $ZA$ associate production or $ZZ$ pair production are transversally
polarized) this pattern can be checked further
experimentally by studying  the distribution of the light
fermions in the $Z \ra
f\bar{f}$ rest frame.

\vspace*{1cm}
\hspace*{9cm}\begin{small}\begin{minipage}[r]{55mm}
{ Fig.~4 Angular distributions of the processes $e^+ e^- \ra ZH,
ZA$ and $ZZ$ at $\sqrt{s}=500$ GeV. The Higgs mass is fixed to $M_H=120$ GeV.}
\end{minipage}\end{small}

\newpage
\nn {\it 2.4.2. WW and ZZ Fusion Processes}

The cross section for the fusion processes,
Eq.~(\ref{wfus},\ref{zfus})  can be written as
\beq
\sigma= \frac{G_F^3 M_V^4}{64 \sqrt{2} \pi^3} \int_{\kappa_H}^1 {\rm d}x
\int_x^1 \frac{ {\rm d}y}{[1+(y-x)/ \kappa_V]^2} \left[ (v^2+a^2)^2 f(x,y)
+ 4 v^2 a^2 g(x,y) \right] \label{fusion}
\eeq
\beq
f(x,y) &=& \left(\frac{2x}{y^3} -\frac{1+2x}{y^2} +\frac{2+x}{2y} -\frac{1}{2}
\right)\left[ \frac{z}{1+z} -\log (1+z) \right] +\frac{x}{y^3}\frac{z^2(1-y)}
{1+z} \non \\
g(x,y) &=& \left(-\frac{x}{y^2}  +\frac{2+x}{2y} -\frac{1}{2} \right)
\left[ \frac{z}{1+z} -\log (1+z) \right] \non
\eeq
with $\kappa_H =M_H^2/s, \kappa_V=M_V^2/s ,z=y(x-\kappa_H)/(\kappa_V x)$ and
$v, a$ the electron couplings to the massive gauge bosons [$v=-1+4s_W^2 ,
a=-1$ for the $Z$ boson and $v=a=\sqrt{2}$ for the W boson].

 Asymptotically it grows as $ M_W^{-2} \log s/ M_H^2$ in contrast to the
processes with s-chaneel exchanges  which fall $\sim s^{-1}$.
The cross section for $ZZ$
fusion is about an order of magnitude smaller than the cross section for $WW$
fusion; this is a mere consequence of the fact that the NC couplings are
smaller than the CC couplings. The lower rate however is, at least partly,
compensated by the clean signature of the $\ee$ final state in
(\ref{zfus})  that allows
for a missing mass analysis to tag the Higgs particle.

For a light Higgs mass, the  dominant background comes from
$\ee \ra e^+W^-\nu_e$ and $ WW \ra Z$ processes.  The cross
sections  exceed the signal for jet--jet final states by about a
factor of 60 and 3, respectively.  Since the single $W(Z)$ boson
production shows a  behavior similar to the Higgs boson in the
signal process, kinematical cuts enhance the signal/background
ratio very little. Only the use of features that are
characteristic: the resonance structure, the spin of the
resonance and the flavor composition of the decays\cite{PR4},
can help in separating signal from background. In particular,
when  the Higgs mass enters the $Z$ and $W$ resonance region,
flavor tagging is indispensable leading to a sample of about 240
events composed of 60 $W \ra jj$, 80 $Z \ra jj$ and 100 $H \ra
jj$ tagged as $b \bar{b}$--jets (again for a luminosity of 20
fb$^{-1}$ and for realistic tagging efficiencies). For a Higgs
mass around $2M_W$, the background process with
$W^+W^-$ and $WZ$ final states can be reduced to a negligible level.

\vspace{3mm}
\nn  {\it 2.4.3. Radiation off the Top}

The coupling of the Higgs boson to other particles is proportional to
their masses. Therefore for a heavy top quark the Higgs radiation off
the top quark-antiquark pair produced in $\epem$ collisions
becomes an interesting process to look at.
 In general, the top and
Higgs masses must be kept non--zero in the calculations so that
the cross section for Higgs
bremsstrahlung is quite involved. It can be found in Ref. \cite{X8,X11}
The integrated cross section is shown for various c.m. energy values
in Fig.~5 as a function of $M_H$. At $\sqrt{s}=500$ GeV, while for small
$M_H$ the cross sections increase with $m_t$ as a result of the rising Yukawa
coupling, this trend is reversed for heavy	Higgses by

\newpage
\vspace*{1cm}
\nn \hspace*{9cm}\begin{small}\begin{minipage}[r]{60mm}
 { Fig.~5 The cross sections $\sigma(e^+e^- \rightarrow t \bar{t}H)$
at $\sqrt{s}=0.5, 1$ and $1.5$ TeV as a function of the Higgs mass; the
top quark mass is fixed to 175 GeV.} \end{minipage}\end{small}

\vspace{25mm}
\nn the reduction of the available phase space. For an
integrated luminosity of $\int{\cal{L}}	= 20 \mbox{ fb}^{-1}$, some
100 events can be expected at Higgs masses of order 60 GeV, falling to
less than 20 events at 100 GeV.
With such a small sample of events the process $\ee \ra
t\bar{t}H$  is not suitable for Higgs
search. However an interesting point is that
the $t\bar{t}H$ final state is generated almost {\it exclusively} through Higgs
bremsstrahlung off the top quarks. There is an additional
contribution with Higgs
particles emitted by the $Z$ line followed by $Z\ra t\bar{t}$
which turns out to be small, of the order 1 -- 2 \%.
Therefore this process may allow for a {\it direct} measurement
of Yukawa coupling of the Higgs to top quarks.
There is a reasonable hope to isolate these events experimentally
despite the low rates since the signature of
the process $e^+e^- \ra t\overline{t}H \ra WWbbbb$	is
spectacular. The large number of $b$ quarks together
with the mass constraints
$m(bb) = m(H)$ and $m(Wb) = m(t)$ will be crucial in rejecting background
events.

Even though Higgs bremsstrahlung off top quarks is not	easy to
handle experimentally
in view	of  small cross sections, it nevertheless deserves attention
as it	may provide the	opportunity to measure directly the
Higgs--fermion coupling.

\subsection{Higgs production at future $\ga$ colliders}
\vspace{-0.35cm}
Since the photons couple to Higgs bosons via
heavy particle loops, the $H\ga$ amplitudes are sensitive to all
charged particles, standard and nonstandard, well above the
Higgs masses themselves.
The cross section for the $\ga$ fusion of Higgs bosons is found by folding the
parton cross section with the $\ga$ luminosity, see
e.g.~Ref.\cite{Zerwas,chiap} and
the parton cross section is determined by the $\ga$ width ($s_{\ga}$ is the
photon c.m. energy squared),
\beq
\sigma( \ga \ra \Phi ) = \frac{8 \pi^2}{M_{\Phi}^3} \Gamma(\Phi \ra \ga) \delta
(1- M_\Phi^2/s_{\ga}) \label{ggfusion}
\eeq

With increasing energy of $\ee$ collider the $\ga$ luminosity of
bremsstrahlung photons increases and $\ga$ fusion may become an
important production process. For example, assuming integrated
luminosity of 10 fb$^{-1}$ at $\sqrt{s}=2$ TeV $\ee$ collider (CLIC) it was
found\cite{chiap} that  some 20 - 30 events
can be expected  for $M_H$ in the range 50 - 250 GeV, falling
quickly to  1  at 400 GeV.   Here relatively low rates are due to the
fact that for bremsstrahlung photons the luminosity is still low
and falls very quickly with the energy available in $\ga$
collisions.

The main interest in $\ga$ fusion, however, stems from the fact that
future high--energy $\ee$ linear colliders  can be {\it converted} to
a high energy $\ga$ colliders,\cite{PPC}
by using Compton back--scattered laser light. In this way one
converts electrons and positrons to energetic photons  with
{\it practically} the same total energy
($\sim 80$\%) and luminosity $(\sim 100$\%) as the the original $\ee$
collider.

Studies performed in Ref.\cite{RichardJikia} show that in a $\ga$
collider one can expect as many as $10^4$ events per year.
Unfortunalely there will be a copious production of
$b\bar{b}$ and $WW$  pairs $\sim 10^5 - 10^6$/year as well which
will complicate Higgs search. However once the Higgs boson is
found and its mass measured,  detailed studies of its properties can
be undertaken by tuning the  energy of $\ga$ collider
to $M_H$. Operating with photons of given polarization may help
to suppress the background.

In particular, we will see in section 3 that in the $\ee$
mode, the lightest
supersymmetric Higgs particle $h$ can be detected but it cannot be
distinguished from the $\SM$ Higgs boson in some part of the $\MSSM$ parameter
space if $\SUSY$ decays are not allowed. One of  the ways to
distinguish $h$ from the $\SM$ Higgs particle can be  provided
by Higgs production
in $\gamma \gamma$ fusion. In the Standard Model this process is
built up by $W$ and top quark loops which interfere
destructively. Additional contributions,
(provided for example by supersymmetric particles: chargino, sfermion and
charged Higgs boson loops) can alter the $\SM$ production rates
and thus help in revealing the nature of Higgs boson.

\section{Search for the Higgs bosons in supersymmetric extension
of the Standard Model}
\vspace{-0.7cm}
\subsection{The Higgs sector of the $\MSSM$}
\vspace{-0.35cm}
 The minimal supersymmetric extension $\MSSM$\cite{Gunion} employs only two
doublets $\Phi_{1}$ and $\Phi_{2}$.
The field $\Phi_{2}$  couples only to
up--type quarks while $\Phi_{1}$ couples to down--type quarks
and charged leptons. The physical Higgs bosons
introduced in this extension are of the following type: two ${\cal CP}$--even
neutral bosons $h$ and $H$ (where by convention $M_h\leq M_H$), a ${\cal
CP}$--odd neutral boson $A$ (usually called pseudoscalar) and
two charged Higgs bosons $H^{\pm}$.

The properties of the scalar particles and
their interactions with gauge bosons and fermions depend on
Higgs masses, $M_h$, $M_H$, $M_A$ and $M_{H^\pm}$, and two
mixing angles: $\alpha$ in the ${\cal CP}$--even neutral,  and
$\beta$ in the charged Higgs sectors.
The angle $\beta$ is related to 	the ratio of the
vacuum expectation values $\tb = v_2/v_1$ (where $v_1$ ($v_2$) is
the vaccum expectation value  of the field $\Phi_1$ ($\Phi_2$)).
 However supersymmetry leads to several relations
among these parameters and, in fact, only two of them are independent at tree
level.  These relations impose a strong hierarchical structure on the mass
spectrum $ ( M_h\leq M_Z , M_A \leq M_H$ and $M_W \leq
M_{H^\pm})$  which however is
broken by radiative corrections\cite{S5} since the top quark mass is large. The
parameter $\tg$ will  be assumed in the range $1 < \tg <
m_t/m_b$, consistent with restrictions that follow
from interpreting the $\MSSM$ as a low energy limit of a supergravity model.

Since the lightest ${\cal CP}$--even scalar boson $h$ is likely
to be discovered first, an attractive choice of the two input
parameters is the set $(M_h, \tb $). Once these two parameters
(and the top quark
mass and the associated squark masses which enter through radiative
corrections) are specified, all other quantities are
predicted.  The most important radiative corrections\cite{S5} can be
determined by the parameter $\epsilon$
\begin{eqnarray}
\epsilon = \frac{3 \alpha}{2 \pi} \frac{1}{s_W^2 c_W^2}
\frac{1}{\sin^2	\beta} \frac{m_t^4}{M_Z^2} \log	\left( 1+
\frac{M_S^2}{m_t^2} \right) \label{eps}
\end{eqnarray}
where $s_W^2=1-c_W^2 \equiv \sin^2 \theta_W$.  These corrections
shift the mass of	the light neutral Higgs boson $h$ upward with
increasin top mass.	The upper limit on $M_h$ as a function
of top quark mass is shown	in Fig.~6 for  squark masses
$M_S=$ 1 TeV and two representative values of $\tb =2.5$ and 20;
and update of Ref.\cite{X5}   The upper
bound on $M_h$ is shifted from the tree level value
$M_Z$ up to $\sim $ 130 GeV for $m_t=175$ GeV and $\sim$ 140 GeV for $m_t=200$
GeV.

\vspace*{1cm}
\nn \hspace*{11cm}\begin{small}\begin{minipage}[r]{40mm}
 { Fig.~6 The masses of the Higgs particles in the $\MSSM$ including
radiative corrections. a) Upper
limit on $M_h$ as a function of $m_t$; b)--d) masses of the $H,A$ and $H^\pm$
Higgs bosons as functions of $M_h$ with the top mass fixed to 175
GeV. The dashed curve
shows the leading correction, $A_t=A_b=\mu=0$, while the solid curves include
the full corrections,  $A_t=A_b=1$~TeV and $\mu=-200,0,200$~GeV;
$\mu$ is a $\SUSY$ Higgs mass and $A_t$, $A_b$ are soft $\SUSY$
breaking couplings.   }
\end{minipage}\end{small}

\newpage
 It is interesting to note that for $m_t=175$ GeV the upper limit
for $M_h$ ($\leq 135$ GeV) is smaller than the lower limit of
$\SM$ Higgs mass  ($\geq 160$ GeV) --
see Fig.~1. Therefore the measurement of $M_h$ might give the
first indication which model is realized in Nature.

Taking $M_h$ and $\tb$ as the input parameters,  masses of other
Higgses are given by
\begin{eqnarray}
M_A^2&= &\frac{M_h^2(M_Z^2-M_h^2+\epsilon)-\epsilon M_Z^2 \cos^2 \beta}
{M_Z^2 \cos^2 2\beta -M_h^2+ \epsilon \sin^2 \beta} \non \\
M_H^2 &	= & M_A^2+M_Z^2-M_h^2+\epsilon	\non \\
M_{H^\pm}^2 & =	& M_{A}^{2}+M_{W}^{2}
\end{eqnarray}
In the subsequent discussion, we will use for definiteness the  values
$m_t=175$ GeV and $M_S=1$ TeV. For the two representative values of $\tb$
introduced above, the masses $M_A,M_H$ and $M_{H^\pm}$ are displayed 	in
Fig.~9b--d as a function of the light neutral Higgs mass $M_h$.
Notice that for  $M_h < M_h^{max}$ for a given value of
$\tb$, the Higgs masses are of the order  100 to 200 GeV for
$M_H$ and $M_{H^\pm}$, and up to $\sim	$ 150 GeV for $M_A$ (similarly
to $M_h$). On general grounds, the masses of the heavy neutral and charged
Higgs bosons are expected to be of the order of the electroweak symmetry
breaking scale.

The	mixing angle $\alpha$ is expressed as
\begin{eqnarray}
{\rm tg} 2 \alpha = {\rm tg} 2 \beta \frac{M_{A}^{2} +M_{Z}^{2}}{M_{A}^{2}-
M_{Z}^{2}+ \epsilon /\cos 2\beta} \hspace*{2cm} \left[ \ -\frac{\pi}{2} \leq
\alpha \leq 0 \ \right]
\end{eqnarray}

The couplings of Higgs bosons to fermions and gauge
bosons in general depend	on the angles $\alpha$ and $\beta$. Normalized
to the $\SM$ Higgs couplings, they are summarized in Table 1.

\vspace*{0.1cm}
\nn {\small Tab.~1: Higgs bosons couplings in the $\MSSM$ to
up-type and down-type fermions and gauge
bosons (V=W,Z) relative to the $\SM$ Higgs boson couplings.}

\vspace{1mm}
\begin{center}
\begin{tabular}{|c|c|c|c|c|} \hline
$\hspace{0.5cm} \Phi \hspace{0.5cm} $ &$ g_{ \Phi \bar{u} u} $ &
$ g_{\Phi \bar{d}
d} $ & $g_{ \Phi VV} $ \\ \hline
$H_{SM}$ & \ $ \; 1  \;	$ \ & \	$ \; 1  \; $ \	& \ $ \; 1  \; $ \ \\
$h$  & \ $\; \cos\alpha/\sin\beta	\; $ \ & \ $ \;	-\sin\alpha/
\cos\beta \; $ \ & \ $ \; \sin(\beta-\alpha) \;	$ \ \\
 $H$  & \	$\; \sin\alpha/\sin\beta \; $ \	& \ $ \; \cos\alpha/
\cos\beta \; $ \ & \ $ \; \cos(\beta-\alpha) \;	$ \ \\
$A$  & \ $\; 1/ \tg \; $	\ & \ $	\; \tg \; $ \
& \ $ \; 0 \; $	\ \\ \hline
\end{tabular}
\end{center}

\vspace{2mm}
  For $\tg >1$ the couplings	to down	(up)
type fermions are enhanced
(suppressed)	compared to the	$\SM$ Higgs couplings.  If $M_h$ is	very
close to its upper limit for a given value of $\tb$,	the couplings of $h$
to fermions and gauge bosons are $\SM$ like.  It may therefore
be very difficult to
distinguish the Higgs	sector of the $\MSSM$ from the $\SM$, if all other
Higgs bosons are very heavy\cite{Haber}.

\vspace{3mm}
\nn {\it Comment on LEP 1 results for $\MSSM$  and $\THDM$ Higgs bosons}

As we already noticed, in $\MSSM$  there are only two independent
parameters in the Higgs sector, for example $M_h$ and $\tb$.
At LEP 1 energies the heavier $\CP$ Higgs boson cannot be
produced on-shell ($M_H>M_Z$) and there are only two important production
mechanisms for $h$ and/or $A$:
$Z \rightarrow Z^*h$ and $Z \rightarrow Ah$.
Negative results of $\SUSY$ Higgs particles searches at LEP in
these processes  exclude\cite{LEPSUSY}  $h$ and $A$
bosons with masses smaller than $M_{h,A} \simeq 45$ GeV for $m_t=140$ GeV,
$M_S=1$ TeV and tg$\beta >1$. If the parameters of the model are allowed to
vary arbitrarily and if one includes all possible decay modes, this bound
becomes\cite{LEPSUSY} $M_h >44$ GeV and $M_A>21$ GeV.

In a general two-Higgs doublet model the  Higgs mass regions
excluded experimentally are much more model dependent.
The reason is that without supersymmetry the masses
of Higgs bosons ($M_h$, $M_H$, $M_A$ and $M_{H^{\pm}}$)  and mixing angles
($\alpha$ and $\beta$) are arbitrary and unrelated parameters. Therefore the
other $\CP$-even Higgs boson $H$ can be produced if light
enough. In addition other production processes, namely Higgs
($h$, $H$ or $A$) radiation off the $b$-quark or $\tau$-lepton
pairs from $Z$ decays can be important\cite{bbh}. For example for large
$\tb$ the couplings of $h$ and $A$ are enhanced to down-type
fermions. Therefore if $ZZh$ is
supressed by mixing angles and the   Higgs pair $Ah$ production
is supressed kinematically then the  radiation of $h$ or $A$
(whichever is lighter) off $b$-quarks or $\tau$-leptons becomes
dominant\cite{Nilles}. (In $SUSY$ model for large $\tb$ and $h$ light enough
to be produced the
$CP$-odd $A$ is almost mass degenerate with $h$ so Higgs pair $hA$ production
process is kinematically possible and dominates.) This is
illustrated in Fig.~7, where
 contour lines for cross sections of processes (a) $\ee\ra hZ\ra
hb\bar{b}$ and
(b) $\ee \ra b\bar{b} \ra hb\bar{b}$ are shown as functions of $M_h$ and $\tb$
and for a specific choice of other parameters.  Experimentally
the radiation processes, which have a different signature, have
not been looked for.  Before these processes are analysed we
cannot conclude that LEP 1 excludes $\THDM$ Higgs bosons within
kinematical reach in a model independent way.

\vspace{1cm}
\nn \hspace*{9.5cm}\begin{small}\begin{minipage}[r]{55mm}
{ Fig.~7 Regions in the plane ($M_h$, $\tb$=$v_2/v_1$)
corresponding to the dominant  process (a) or (b). Contour lines are
drawn for log$_{10}(\sigma/1$pb). Solid (dashed) lines are for
$\epsilon=0.2 (2)$. For details, see Ref.\cite{Nilles}}
\end{minipage}\end{small}

\newpage
\subsection{Decay modes }
\vspace{-0.35cm}
\nn {\it 3.2.1.  Decays to $\SM$ particles}

The partial decay width of a neutral Higgs boson $\Phi$ into fermion
pairs is given by
\begin{eqnarray}
\Gamma(\Phi \ra \bar{f} f) & = & N_{c} \frac{G_F m_f^2}{4\sqrt{2} \pi}
\ g_{\Phi ff}^2 \ M_{\Phi} \ \beta^{p}
\end{eqnarray}
where $\beta=(1-4m_f^2/M_{\Phi}^2)^{1/2}$ and  $p$ = 3(1) for the ${\cal
CP}$--even (odd) Higgs boson; the couplings $g_{\Phi ff}$ are listed in Tab.1.
For final state quarks one has to include QCD corrections. As in
$\SM$ the use of the running
masses takes into account the bulk of these corrections, which in the
limit $M_H \gg m_q$ are the same for $\CP$--odd and $\CP$--even
Higgs bosons.

The lightest Higgs boson will decay mainly into fermion pairs since its
mass is smaller than $\sim$ 130 GeV. This is also the dominant decay mode
of the pseudoscalar boson $A$ which has no tree level couplings to gauge
bosons.
For values of $\tb$ larger than unity and for masses less than $\sim$ 130 GeV,
the	main decay modes of $h$ and $A$ will be decays into $b
\bar{b}$ and $\tau^+ \tau^-$ pairs with the branching
ratios being always
larger than $ \sim 90\%$ and $5\%$, respectively. The decays into $c \bar{c}$
and gluons are in general strongly suppressed especially for large values of
$\tb$. For large masses, the top decay channels $H,A \rightarrow t\bar{t}$ open
up; yet this mode remains suppressed for large $\tb$.

If the mass is high enough, the heavy ${\cal CP}$--even Higgs boson can in
principle decay into weak gauge bosons $H \rightarrow VV$, $V = W$ or $Z$.
Below the threshold it can decay into
$VV^*$ pairs with  one of the vector bosons being
virtual. (For $h$ this decay mode is negligible.) The partial
decay width is  given by
\begin{eqnarray}
\Gamma (H \ra VV^{(*)}) =  g_{HVV}^2 \Gamma (H_{\rm SM} \ra VV^{(*)})
\end{eqnarray}
where $\Gamma (H_{\rm SM} \ra VV^{(*)})$ given by
Eqs.~(8,9,10).  Since the $H$ partial width is proportional
to $\cos^2 (\beta-\alpha)$, it is strongly suppressed because if
$M_H$ is large
enough for these decay modes to be kinematically allowed, $M_h$ is very close
to its maximum so that $\cos^2 (\beta -\alpha) \rightarrow 0$. For the same
reason, the decay $ A
\rightarrow	Zh$, is also suppressed\cite{Gunion}
\begin{eqnarray}
\Gamma(A \ra Zh) = \frac{G_F}{8\sqrt{2} \pi} \cos^2 (\beta-\alpha)
\ \frac{M_Z^4}{M_A} \lambda^{1/2}(M_Z^2,M_h^2;M_A^2) \lambda(M_A^2,M_h^2;M_Z^2)
\end{eqnarray}
\nn with $\lambda(x,y;z)=(1-x/z-y/z)^2-4xy/z^2$ being the usual two--body phase
space function.

The heavy neutral Higgs boson $H$ can also decay into two lighter Higgs
bosons $h$ or $A$,\cite{Gunion}
\begin{small}
\begin{eqnarray}
\Gamma(H \ra hh) &=& \frac{G_F}{16\sqrt{2} \pi} \frac{M_Z^4}{M_H}
\left(1-4\frac{M_h^2}{M_H^2} \right)^{1/2} \left[ \cos 2\alpha \cos(\beta+
\alpha)-2 \sin 2\alpha \sin (\beta+\alpha)\right]^2 \nonumber \\
\Gamma(H \ra AA) &=& \frac{G_F}{16\sqrt{2} \pi} \frac{M_Z^4}{M_H}
\left(1-4\frac{M_A^2}{M_H^2} \right)^{1/2} \left[ \cos 2\beta \cos(\beta+
\alpha) \right]^2
\end{eqnarray}
\end{small}
\nn These modes, however, are restricted to small domains in
the parameter space.

Decays $\Phi \ra gg$ are mediated by top and bottom quark
loops since the squarks decouple from the effective $\Phi gg$ vertex for high
masses. This decay mode is small. The branching ratio
reaches a few \%  for $h$-decay only for
$M_h$ close to the maximum value where $h$ has
$\SM$ like couplings, and for $H$-decay only below 140 GeV and small
values of $\tb$ where the coupling
to top quarks is sufficiently large.  For the pseudoscalar
Higgs particle, the gluonic decay mode is also small.
Decays of Higgs bosons to $\gamma
\gamma$ and $Z \gamma$ final states are very rare with branching ratios of
order ${\cal O}(10^{-3})$ or below.

The	coupling of the charged Higgs particle to fermions is a
mixture of scalar and pseudoscalar couplings
\begin{eqnarray}
g_{H^+u\bar{d} } = \left( \frac{G_F}{\sqrt{2}}\right)^{1/2} \left[(1-\gamma_5)
 \frac{m_u}{\tb} + (1+\gamma_5) m_d \tb \right]
\end{eqnarray}

The charged Higgs particles decay into fermions with a partial decay width
\beq
\Gamma(H^+ \rightarrow u \bar{d}) =   \frac{N_c G_F \lambda^{\frac
{1}{2}}}
{4 \sqrt{2} \pi M_{H^\pm}}
 \left[(M_{H^\pm}^{2} -m_{u}^{2}-
m_{d}^{2})
 \left( m_{d}^{2} {\rm tg}^2 \beta + \frac{m_u^2} {{\rm tg}^2 \beta}
\right) -4m_u^2m_d^2 \right]
\eeq
\nn with $\lambda^{\frac{1}{2}}=\lambda^{\frac{1}{2}}(m_u^2,m_d^2;
M_{H^{\pm}}^2)$
and, if allowed kinematically, they also decay into the lightest
neutral Higgs plus a $W$ boson,
\begin{small}
\begin{eqnarray}
\Gamma(H^{+} \rightarrow Wh) = 	\frac{G_F \cos^2 (\beta-\alpha) }{8\sqrt{2}\pi
c_W^2} \frac{M_W^4}{M_{H^\pm}} \lambda^{\frac{1}{2}}(M_W^2,M_H^2;M^2_{H^{\pm}})
\lambda(M_{H^\pm}^2, M_h^2; M_W^2) \hspace*{0.2cm}
\end{eqnarray}
\end{small}

\vspace{1cm}
\nn \hspace*{9cm}\begin{small}\begin{minipage}[r]{60mm}
 { Fig.~8 Total decay widths of the $\SUSY$ Higgs bosons (without
decays into $\SUSY$ particles) as functions of their masses for (a) ${\rm tg}
\beta=2.5$ and (b) ${\rm tg}\beta =20$. The top mass was chosen as $m_t=175$
GeV and $M_S=1$ TeV.}
\end{minipage}\end{small}

\newpage
Below the $tb$ and $Wh$ thresholds, the charged Higgs particles will decay
mostly into $\tau \nu_\tau$ and $c\bar{s}$ pairs, the former being dominant for
$\tb >1$.  For large $M_{H^\pm}$ values, the top--bottom decay $H^+
\rightarrow t\bar{b}$ becomes dominant.

Adding up the various decay modes,\cite{X5,PR3} the width of all five
Higgs bosons remains very small, even for large masses. This is shown for the
two representative values $\tb=2.5$ and 20 in Fig.~8. Apart from the ${\cal
CP}$--even heavy neutral Higgs boson $H$ and small $\tb$, the pattern of
branching ratios is in general quite simple. The neutral Higgs bosons decay
preferentially to $b \bar{b}$, and to a lesser extent to $\tau^+ \tau^-$ pairs;
the charged Higgs bosons to $\tau \nu_\tau$ and, preferentially, $t \bar{b}$
pairs above this threshold.

\vspace{3mm}
\nn {\it 3.2.2.  Decays to supersymmetric particles}

At least for the heavy Higgs bosons
$H,A$ and $H^\pm$ decays into charginos and neutralinos could
eventually play a significant role
since some of these particles are expected to be light enough.
 From the negative search of supersymmetric particles in $Z$ decays,\cite{R8}
the lightest neutralino [$\tilde{\chi}_1^0$] mass is restricted to be larger
than 20 GeV for $\tb=2.5$ and larger than 22 GeV for $\tb >4$; the second
lightest neutralino [$\tilde{\chi }_2^0$] and the charginos are excluded if
their masses are less than $ \sim M_Z/2$. If the search at LEP200 with a c.m.
energy of 180 GeV is negative, charginos with masses $m_{\tilde{\chi}^+_1}<90$
GeV will also be excluded.

\vspace{10cm}
\nn {\small Fig.~9. Contour lines in the $(\mu,M$) plane where the sum of the
branching ratios of the lightest Higgs boson $h$ into charginos and neutralinos
exceeds 5\% (dashed lines) and 50\% (full lines) for two values of $M_h$
and $\tb$; the shaded areas are the regions which are (can be) excluded at
LEP 1 (LEP200). The top mass was taken to be $m_t=140$ GeV.}

\newpage
Sfermions are probably too heavy to
affect Higgs decays. The present LEP data exclude sleptons with
masses below $\sim M_Z/2$; if sleptons will not be observed at LEP200  these
limits can be improved by roughly a factor of two. On the other hand, CDF
data\cite{CDFSQUARK} restrict the squarks masses to be larger than $\sim$ 150
GeV if cascade decays are suppressed.  We shall assume in this
discussion that squarks and sleptons are heavy so that they will not affect
Higgs boson decays and production.

The decay widths of the neutral and charged Higgs bosons into chargino or
neutralino pairs  can be found in Ref.\cite{V12}
In Fig.~9 (an update of  Ref.\cite{X5,inv}) the
contour lines  are shown in the ($\mu,M$) plane where the sum of the
branching
ratios of $h$ into the lightest chargino pair and the lightest and
next--to--lightest neutralino pairs exceeds 5\% (dashed lines)
and 50\% (solid lines). ($M$ is a universal gaugino mass.)
These decays can be important for $h$ masses close to
the maximum allowed values; in this case the lightest Higgs boson
has $\SM$ couplings and the dominant $b\bar{b}$ decay mode is not enhanced
anymore for large $\tb$ values, so that other decay modes can become
significant. For these masses and for large $\tb$ values, the branching ratios
for neutralino/chargino decays are sizable even outside the regions which can
be probed at LEP200.

The decays of $H$, $A$ or $H^{\pm}$ into chargino and neutralino pairs can be
very large.\cite{X5} They can exeed 50\% in some areas of
the $\MSSM$ parameter
space, $i.e.$  for positive values of $\mu$ and/or $M$ values below 200
GeV. This is due to the fact that the couplings of the Higgs
bosons to charginos and neutralinos are gauge couplings which can be larger
than the couplings to standard fermions and  gauge
bosons.  Finally, the
branching fractions of the invisible neutral Higgs decays can be important,
\cite{Abdel,X5}  and they could jeopardize the search for the Higgs
particles at hadron colliders. However, at $\epem$ colliders
missing mass techniques can be used to detect invisibly decaying
Higgs particle produced in association with the $Z$ boson or in
mixed visible and invisible decays modes in $Ah$ or $HA$ production.

\subsection{Production mechanisms of $\MSSM$ Higgs bosons at
future $\ee$ colliders}
\nn {\it 3.3.1. Neutral Higgs Bosons}

The main production mechanisms of neutral Higgs bosons at $\ee$
colliders are the bremsstrahlung,  pair production and fusion processes
\begin{eqnarray}
i) \ \ {\rm bremsstrahlung} \hspace{1cm} \ee & \ra &  (Z) \ra Z+h/H
\hspace{4cm} \\
ii) \ \ {\rm pair \ production} \hspace{1cm} \ee & \ra & (Z) \ra A+h/H \\
iii) \ \ {\rm fusion \ processes} \hspace{0.8cm} \ \ee & \ra &  \nu \ \bar{\nu}
\ (WW) \ra \nu \ \bar{\nu} \ + h/H \hspace{2.3cm} \nonumber  \\
\ee & \ra &  \ee (ZZ) \ra \ee + h/H
\end{eqnarray}
The ${\cal CP}$--odd Higgs boson $A$ cannot be produced in fusion processes
to leading order. The Higgs radiation off top quark is supressed
for $\tb >1$.

\newpage
The cross sections for the production
processes $i)$ and $ii$) can be written as
\begin{eqnarray}
\sigma(\ee \ra Zh) & =& \sin^2(\beta-\alpha) \ \sigma_{\rm SM} \nonumber \\
\sigma(\ee \ra ZH) & =& \cos^2(\beta-\alpha) \ \sigma_{\rm SM} \nonumber \\
\sigma(\ee \ra Ah) & =& \cos^2(\beta-\alpha) \ \sigma_{\rm SM} \ \bar{\lambda}
\nonumber \\
\sigma(\ee \ra AH) & =& \sin^2(\beta-\alpha) \ \sigma_{\rm SM} \ \bar{\lambda}
\label{sigmas}
\end{eqnarray}
\nn where
$\sigma_{\rm SM}$ denotes the  cross section for Higgs
bremsstrahlung in $\SM$, Eq.~(\ref{sigmaZH}), and
\begin{eqnarray}
\overline{\lambda} \ = \ \frac{\lambda^{3/2}(M_{j}^2,M_A^2; s)}{\lambda^{1/2}
(M_{j}^2,M_Z^2; s) \ (\lambda(M_{j}^2,M_A^2; s)+12M_Z^2/s) } \hspace*{1cm}
( \ j=h,H \ )
\end{eqnarray}

\vspace{1cm}
\nn \hspace*{10.5cm}\begin{small}\begin{minipage}[r]{45mm}
{ Fig.~10 Production cross sections of the ${\cal CP}$--even neutral
Higgs bosons at $\sqrt{s}=500$ GeV as functions of their masses for three
values of ${\rm tg}\beta=2.5,5$ and 20; (a) Bremsstrahl processes $e^+e^-
\rightarrow Z+h/H$, and (b) in association with the pseudoscalar Higgs boson
$e^+e^- \rightarrow A+h/H$.}
\end{minipage}\end{small}

\newpage
\nn accounts for the correct suppression of the $P$--wave
cross sections near the threshold.
 Since the cross sections in Eq.~(\ref{sigmas}) come with either
with a coefficient $\sin^2(\beta-\alpha)$
or $\cos^2(\beta-\alpha)$ they  are mutually complementary
to each other. Therefore at least one of the
${\cal CP}$--even Higgs bosons should be detected because $\sigma_{\rm
SM}$ is large. Let us discuss them in more detail.

 The cross sections for the production of the
bosons $h$ and $H$ via bremsstrahlung are shown as functions of the
Higgs mass in Fig.~10a. The cross section for $h$ is large for small values of
$\tb$ and/or large values of $M_h$ where $\sin^2(\beta- \alpha)$ approaches
maximum value.	In both cases the cross section is	of the order of
$\sim 50$ fb, which  for an integrated luminosity of 20 fb$^{-1}$ corresponds
to $\sim$ 1000 events.	On the other hand, the cross section for $H$ is
large for large $\tb$ and light $h$ (implying small $M_H$). In the case of
$h$ (and also for $H$ in most of the parameter space) the signal consists
of a $Z$ boson accompanied by a $b\bar{b}$ or a $\tau^+ \tau^-$ pair.
The signal is easy to separate from the background which comes mainly
from $ZZ$ production if the Higgs mass is close to $M_Z$.
Results for the associate production $\ee
\rightarrow Ah$ and $AH$
are displayed in Fig.~10b. The situation is opposite to the
previous case: for light $h$ and/or
large values of $\tb$ the cross section for $Ah$ is large
whereas $AH$ production is preferred in the complementary
region. The sum of the two cross sections decreases from $\sim 50$ to 10 fb
if $M_A$ increases from $\sim 50$ to 200 GeV. Signals mainly
consist of four $b$ quarks in the final state, requiring
efficient $b$ quark tagging. Mass constraints will help to
eliminate the backgrounds from QCD jets as well as $ZZ$ final states.
The above discussion  is summarized in Fig.~11, from
Ref.\cite{Abdel}, an update of Ref.\cite{X5}

\vspace{1cm}
\nn \hspace*{9cm}\begin{small}\begin{minipage}[r]{60mm}
{ Fig.~11 Regions of the $(M_h,\tb)$ plane where
$\ee \ra hZ, hA, HZ$ and $HA$ are observable, $i.e.$
the cross sections are larger than 2.5 fb. The dashed
area is  theoretically forbidden. The region to
the left of the thin line  can be probed at
LEP200. The process $\ee \ra hZ$ is accessible in the
entire area below the full line, $hA$ in the entire area
above the broken line and $HZ$ in the entire area above the full
line; $HA$ final states can be detected in the area between the two
dashed lines.}
\end{minipage}\end{small}

\newpage
 $WW$ and $ZZ$ fusion provide additional mechanisms for the production
of the ${\cal CP}$--even neutral Higgs bosons. They
can again be expressed in terms of the corresponding $\SM$ cross
sections, given in Eq.~(\ref{fusion})
\begin{eqnarray}
\sigma( \ee \ra (VV) \ra h) & = & \sin^2 (\beta-\alpha) \sigma_{\rm SM}^{VV}
\nonumber \\
\sigma( \ee \ra (VV) \ra H) & = & \cos^2 (\beta-\alpha) \sigma_{\rm SM}^{VV}
\end{eqnarray}
\nn As in
the case of the bremsstrahlung process, the production of light
$h$ and heavy $H$ bosons are  complementary.

\vspace*{1cm}
\nn \hspace*{10cm}\begin{small}\begin{minipage}[r]{50mm}
{ Fig.~12 Production cross sections of the ${\cal CP}$--even neutral
Higgs bosons at $\sqrt{s}=500$ GeV as functions of their masses for three
values of ${\rm tg}\beta=2.5,5$ and 20; (a) $WW$ fusion process $e^+e^- \ra
\nu \bar{\nu} +h/H$ (b) and  $ZZ$ fusion process $e^+e^- \ra e^+e^-+h/H$.}
\end{minipage}\end{small}

\vspace{8.7cm}
For the Higgs mass less  than 160 GeV at $\sqrt{s} = 500$ GeV  the
$WW$ fusion mechanism has cross sections larger than
the bremsstrahlung process. However, the final state cannot be fully
reconstructed and the signal is more difficult to extract.  The
cross sections for the $ZZ$ fusion mechanism are about an
order of magnitude smaller than for the one for $WW$.
Nevertheless it will be useful as the final state can
be fully reconstructed. For representative values of $\tb$ they
are shown in Fig.~12.

\newpage
\nn  {\it 3.3.2. Charged Higgs Bosons}

An unambiguous signal of an extended Higgs sector would	be the
discovery of a charged Higgs boson.  In	a general two--Higgs doublet
model, charged	Higgs bosons can be as	light as $\sim $ 45 GeV, the
lower limit derived from the negative search at LEP 1.\cite{R8}  In the
$\MSSM$ however, $H^\pm$ is constrained to be heavier than the $W$ boson.
More precisely, the lower limit $M_A >45$ GeV obtained at LEP 1
implies $M_{H^\pm} > 90$ GeV.

The production of a pair of charged Higgs
bosons in $\ee$ collisions proceeds through virtual
photon and $Z$ boson exchange.
\begin{eqnarray}
\sigma(	\ee \longrightarrow H^{+}H^{-}) = \frac{\pi \alpha^2}{3s}
\left[ 1- \frac{ 2 \hat{v}_e \hat{v}_H}{1-M_Z^2/s} + \frac{(\hat{a}_e^2
+\hat{v}_e^2)\hat{v}_H^2} {(1-M_Z^2/s)^2}  \right] \ \beta^3
\end{eqnarray}
with the standard $Z$ charges $\hat{v}_e=(-1+4s_W^2)/4c_Ws_W$, $\hat{a}_e
=-1/4c_Ws_W$ and $\hat{v}_H=(-1+2s_W^2)/2c_Ws_W$, and $\beta=(1-4M_{H^\pm}^2/s
)^{1/2}$.
The cross section is shown in Fig.~13a as a function of the charged Higgs mass
for a c.m. energy $\sqrt{s}= 500$ GeV (it does not depend on
any extra parameter). For small Higgs masses the cross section
is of order 100 fb, but it drops very quickly due to the P--wave suppression
factor $\beta^3$ near the threshold. For $M_{H^\pm}=$ 220 GeV, the cross
section has fallen to a level of $ \simeq $ 5 fb, which for an integrated
luminosity of 20 fb$^{-1}$ corresponds to 100 events.  The
angular	distribution of the charged Higgs bosons follows the $\sin^2 \theta$
law typical for spin--zero particle production.

\vspace*{1cm}
\nn \hspace*{10cm}\begin{small}\begin{minipage}[r]{50mm}
{ Fig.~13 Production cross sections of the charged Higgs boson (a)
at $e^+ e^-$ colliders with $\sqrt{s}=500$ GeV and in $\gamma \gamma$
collisions
with $\sqrt{s}=400$ and (b) in top decays with $m_t=175$ GeV and for three
values of $\tb=2.5,5$ and 20.}
\end{minipage}\end{small}

\newpage
The charged Higgs boson, if lighter	than the top quark, can also be
produced in top decays.\cite{W6b}
In the range $ 1 < \tb < m_t/m_b$ favored by $\SUSY$ models, the  branching
ratio BR$(t \rightarrow bH^+)$ varies between $ \sim 2 \%$ and $20 \%$.  Since
the cross section for top pair production is of order of 50 fb at $\sqrt{s}=
500$ GeV, this corresponds to 20 and 200 charged Higgs bosons at a luminosity
$\int {\cal L} =$ 20 fb$^{-1}$; Fig.~13b.

If $M_{H^\pm}
< m_t+m_b$, the charged Higgs boson will decay mainly into $\tau \nu_\tau$
and $c\bar{s} $ pairs, the $\tau \nu_\tau$	mode dominating
for $\tb$ larger than
unity. This results in a surplus of $\tau$ final states over $e, \mu$
final states, an apparent breaking of $\tau \ vs. \ e,\mu$ universality.
For large Higgs masses the dominant decay mode is the top decay $H^+
\rightarrow t \bar{b}$. In some part of the parameter space also the
decay $H^+ \rightarrow  W^+h $ is allowed, leading to cascades with heavy
$\tau$ and $b$ particles in the final state.

\subsection{$\MSSM$ Higgs bosons in $\gamma \gamma$ Collider}
\vspace{-0.35cm}
The scalar $\MSSM$
Higgs particles $h,H$ couple to vector bosons directly, and
therefore the positive parity can
be checked by analyzing the $Z$ final states in $\ee \ra Z^* \ra ZH\ (Z \ra f
\bar{f})$ as discussed above. However, the pseudoscalar $A$ boson has no
tree level couplings to the vector bosons and the latter methods cannot be
used. To study\cite{Wggspin} the ${\cal CP}$ properties of both scalar and
pseudoscalar Higgs bosons on equal footing, one can run in the $\ga$ mode where
both type of particles are produced through loop diagrams with similar rates.
For scalar
particles the production amplitude  is maximal only for parallel
vectors while pseudoscalars prefer perpendicular polarization
vectors.  It remains to be seen whether the degree of
linear polarization of the beams will be high enough to be
efficiently explored.

Charged Higgs particles can	also be	created in $\gamma \gamma$ collisions.
 The cross section is given by
\begin{eqnarray}
\sigma (\gamma \gamma \ra H^+ H^-) = \frac{2\pi \alpha^2}{s}
\beta \left[ 2 -\beta^2 - \frac{1-\beta^4}{2\beta} {\rm log}
\frac{ 1+\beta}{1-\beta} \right]
\end{eqnarray}
where $\beta$ is the velocity of the Higgs particle. The numerical result
is displayed in Fig.~13a  for the $\gamma \gamma$ luminosity without beam
polarization. Due to the reduced energy, the maximum Higgs mass
which can be probed in $\gamma \gamma$ collisions is smaller than in the
original $\ee$ collisions; the cross section however is enhanced by a factor
$\sim 3$ in the low mass range.

\section{Summary}

The Higgs search and tests of the electroweak symmetry
breaking will be the most important physics goals of future
high--energy colliders. In this review, we have discussed the properties of
the Higgs particles of the Standard Model and of its minimal supersymmetric
extension. We have updated the new value of the top quark
various results for Higgs couplings, decay widths and
production cross sections  in  future $\ee$ and $\ga$ colliders.

\newpage
$\ee$ linear colliders with energies in the range 300 -- 500 GeV and a
luminosity of a few times $10^{33}$cm$^{-2}$s$^{-1}$ are ideal
machines to search for Higgs bosons in the mass range below the scale of
electroweak symmetry breaking.

In the intermediate  mass range, Standard Higgs particles can be observed
in three independent production channels:
the bremsstrahlung process $\ee \ra ZH$ and the fusion processes $\ee \ra
\bar{\nu}\nu H$ and $\ee \ra \ee H$. The particle is relatively easy to detect
especially in the $ZH$ channel, where the main background from $ZZ$ pair
production can be suppressed efficiently by using micro--vertex detectors since
Higgs bosons with masses below 140 GeV decay mainly into $b\bar{b}$ pairs.

An even stronger case for $\ee$ colliders operating in the 500 GeV range is
made by supersymmetric extensions of the Standard Model. Since
in $\MSSM$  the lightest Higgs particle has a mass below 140 GeV and decays
mainly into $b\bar{b}$ and $\tau^+ \tau^-$ pairs, it cannot be missed at an
$\ee$ collider with an energy $\sqrt{s} > 300$ GeV, independently of its decay
modes and in the entire $\SUSY$ parameter space. The heavy neutral Higgs
particles can be produced in the bremsstrahlung and fusion processes or
pairwise, these processes being complementary.  At least one neutral Higgs
boson must be detected or $\MSSM$ rejected. In a large part of
the $\SUSY$ parameter space, all
neutral Higgs bosons can be observed. Charged Higgs particles can be detected
up to practically the kinematical limit.

Once
the Higgs boson is found, its fundamental properties can be investigated. The
Higgs spin can be measured by analyzing the angular dependence of the $ZH$
production process and in the Higgs decays into massive gauge bosons. The Higgs
couplings to the massive gauge bosons can be determined through the production
rates, the coupling to heavy fermions through the Higgs decay branching
fractions, and in some mass window, Higgs radiation off top quarks.

Future $e^+e^-$ can be turned to very high--energy $\gamma \gamma$ or $e
\gamma$ colliders by using back--scattering of laser light. The $\gamma \gamma$
mode of the $e^+ e^-$ collider could be useful to measure accurately the
Higgs--photons coupling to which new particles might contribute, and to study
the ${\cal CP}$ properties of the Higgs particles.

Although we did not discuss the hadron colliders in this review,
it turns out that $\ee$ linear colliders operating in the
300--500 GeV energy range and
hadron colliders operating in the multi--TeV range have a complementary
potential\cite{Poggioli} for exploring the key issue of the
mechanism of electroweak symmetry
breaking.

\nonumsection{Acknowledgments}
I would like to thank  Peter Zerwas and Abdel
Djouadi for an enriching, fruitful and enjoyable collaboration
on the Higgs sector. Thanks also go to the organizers of this conference
for the fruitful and pleasant atmosphere they have provided.

\newpage
\nonumsection{References}

\end{document}